\documentclass[runningheads]{llncs}
\usepackage{graphicx}
\usepackage{comment}

\begin{document}
\title{Coupling purposes with status-functions in artificial institutions\thanks{Supported by Federal Institute of Education, Science and Technology of Rio Grande do Sul (IFRS).}}

%\title{A model for coupling purposes in artificial institutions\thanks{Supported by Federal Institute of Education, Science and Technology of Rio Grande do Sul (IFRS).}}
\author{Rafhael R. Cunha\inst{1}\orcidID{0000-0003-3233-5158} \and
Jomi F. Hübner\inst{1}\orcidID{0000-0001-9355-822X} \and
Maiquel de Brito\inst{2}\orcidID{0000-0003-4650-7416}}
\authorrunning{Cunha. Rafhael R, et al.}

\institute{Automation and Systems Department, Federal University of Santa Catarina, Florianópolis, Brazil \\
\email{rafhael.cunha@posgrad.ufsc.br} \\
\email{jomi.hubner@ufsc.br} \and
Control, Automation, and Computation Engineering Department, Federal University of Santa Catarina, Blumenau, Brazil \\
\email{maiquel.b@ufsc.br}}

%\institute{Princeton University, Princeton NJ 08544, USA \and
%Springer Heidelberg, Tiergartenstr. 17, 69121 Heidelberg, Germany
\maketitle              % typeset the header of the contribution
\begin{abstract}
%The abstract should briefly summarize the contents of the paper in 15--250 words.

%In multi-agent systems, artificial institutions connect institutional concepts, belonging to the institutional reality, to the concrete elements that compose the system. In this paper, we focus on the cases where the institutional reality is composed of Status-Functions, as proposed by Searle. Although there are proposals to model Status-Functions and their dynamics from an institutional perspective, we still miss some aspects from an agent perspective.
%
In multi-agent systems, the agents may have goals that depend on a social, shared interpretation about the facts occurring in the system. These are the so-called social goals. Artificial institutions provide such a social interpretation by assigning statuses to the concrete elements that compose the system. These statuses are supposed to enable the assignee element to perform functions that are not exclusively inherent to their design features. However, the enabled functions are not explicit in the existing models of artificial institutions.
As a consequence, (i) agents may have difficulties to reasoning about how to achieve their own social goals with the help of artificial institutions and (ii) these institutions are not well instrumented to receive incoming agents, in the case of open systems.
Considering those problems, this paper proposes a model to express the functions -- or the \emph{purposes} -- associated with the status-functions helping the agents to reason about their social goals and the institution. We evaluate the model by using it in some scenarios, showing how the agents can use purposes to reason about the satisfaction of their social goals in institutional contexts and how the institution can be flexible enough to support new agents operating in the system.

\keywords{purposes  \and status-functions \and artificial institutions \and multi-agent systems.}
\end{abstract}

\section{Introduction}
\label{introduction}

Multi-agent systems (MAS) are systems composed of autonomous computational entities, henceforth referred to as \emph{agents}, that can interact within a dynamic environment to achieve their common and individual goals~\cite{wooldridge2009introduction}. 
The interaction among the agents is in the very core of MAS, making it a useful approach to handle computational problems involving social aspects~\cite{winikoff2012challenges}. One of these aspects is a shared interpretation about the facts occurring in the system.
For example, consider a scenario where (i) the agent \emph{Bob} has the goal of having a book and (ii) the agent \emph{Tom} desires to sell a book. To this end, (i)~\emph{Bob} needs to execute an action that means \emph{giving a value and in return receive a good}, and (ii)~\emph{Tom} waits for such action to then deliver the book. 
In this scenario, both goals are \emph{social goals} because they can not be achieved alone and depend on a \textit{common interpretation} involving certain facts.
Without such common interpretation, \emph{Bob} might not know which action to perform to give a value in exchange for the book. Even this would not be the case, \emph{Tom} might not acknowledge the action of \emph{Bob}, refusing thus to deliver the book. 
The highlighted problem therefore involves social aspects.

Inspired by human societies, some works propose models and tools to provide this kind of interpretation to computer systems and, in particular, to MAS~\cite{Fornara2007}. They usually consider that the elements involved in the interaction among the agents \emph{constitute} (or \emph{count as}) institutional concepts, that are the common interpretation of those concrete elements~\cite{cliffe2006answer,CARDOSO2007,brito2016model,fornara2011specifying}). For example, agents acting in an e-commerce scenario may constitute (or count as) \emph{buyers}, while some of their actions may count as \emph{payments}.
These institutional concepts are referred in the literature as \emph{status-functions}: they are \emph{status} that assign \emph{functions} to the concrete elements~\cite{searle1995construction,searle2010making}. For example, the status \emph{buyer} assigns to an agent some functions such as perform payments, take loans, etc. Artificial Institutions are the component of the MAS that is responsible for defining the conditions for an agent to become a \emph{buyer}, or an action to become a \emph{payment}~\cite{searle1995construction,searle2010making}.

%por mais que SF definem o que conta como o que, elas sozinhas não definem qual é o significado (ou a forma de interpretar) aquela SF. 

%Problema - Descrição clara
The existing works on Artificial Institutions are more concerned with specifying and managing the constitution of status-functions. 
However, they focus more on the status than the function. While the status is explicit, \textit{the function is implicit}.
In other words, the works described in the literature, as far as we know, \emph{do not provide the means for the institution to express the functions associated with the status explicitly}.
The main drawbacks of this limitation are: (i) the agents cannot reason about the functions performed by the elements that carry the status and (ii) the institutions may be incompatible~\footnote{
Compatibility in this work refers to the situation in which the vocabulary used in the specification of the agents works properly with the vocabulary of the institutional specification. Compatibility can occur in several ways, including (i) if the vocabulary present in the agent's specification is identical to the institutional vocabulary or (ii) if the programmer encodes compatibility with the institutional specification within the agent.} with the agents' specification. 
%Consequently, the agents cannot exploit these functions to satisfy their goals and the institutions need to adapt to receive new agents.

%Relevância do Problema

The limitations discussed earlier have some implications. First, different agents may have problems to achieve similar social goals if their goals are internally identified by different terms, even if they are acting in the same environment and institution. It is especially critical in open MAS, where the agents can be designed and implemented by different parties and it is not possible to predict, in design time, neither the number, nor the behaviour, nor the way the agent shall interact among themselves and explore the available resources~\cite{fornara2004agent,Piunti2009}.
For example, in the \emph{book store} scenario, the \emph{payment} status-function can have the function of \textit{leading the system to a state where the agent that executes an action that counts as payment has the book}. In this case, \emph{have a book} is the function of the \emph{payment} status-function interpreted from the agents' perspective. It is also the \emph{Bob's} social goal. Consider that \emph{Alice} joins the \emph{book store} scenario. \emph{Alice} has the social goal of \emph{acquiring a book}. Both Bob and Alice's have similar social goals. However, if the functions associated to the status-functions are not explicit, \emph{Alice} will only reach her social goal if her code includes the goal of \emph{having a book}, similar to \emph{Bob}.
Second, the same agent may not reach his social goal if it moves itself to a new environment and institution similar to the previous with the same goal.
For example, consider a new scenario in the system called a \emph{library}. In it, \emph{have a book}
is the function of a status-function other than \emph{payment}. When entering this system, \emph{Bob} will not achieve his social goal because there is no \emph{payment} status-function there, and he was coded for that very status-function of the \emph{book store}. 
%\emph{Bob} will only reach his goal if he performs the function aligned with his objective associated with another status-function in this institution.
%JH:\emph{Bob} will only achieve his goal if he performs the function associated with him. %that in this system is represented by another status-function.
Then, \emph{Bob} will only reach his social goal if he performs the action with the status-functions associated with his social goal in that institution.
Third, the same institution may not be compatible with different agents with similar social objectives but coded with other terms. This is the same limitation as to the previous example but seen from the institution's perspective. The institution needs to add new status-functions compatible with agents' specifications to its specification to get around this problem. However, some status-functions can make it inconsistent.
For example, it is not possible to specify the \emph{payment} status-function in a library that only lends books.

Although the work on artificial institutions focus on supporting the regulation of the system, institutions also need to help agents through constitutive rules, status-functions, etc. to achieve their social goals~\cite{rodriguez2015towards}.
These examples show that the institutions as they are currently conceived do not specify the functions associated with the status-functions.  
They do not support the reasoning of agents~\cite{rodriguez2015towards} concerning the satisfaction of their own social objectives and are not prepared to receive agents designed by different developers. Thus, the main contribution of this paper is a model based on the notion of \emph{purpose} to explicitly represent the functions of the status-functions and to relate them to the social goals of the agents. 
It is inspired by the ``Construction of the Social Reality'' by John Searle~\cite{searle1995construction,searle2010making} and ``Documentality'' by Maurizio Ferraris~\cite{condello2019money} both philosophers' theories that seem to be fundamental for comprehend the social reality.

This paper is organized as follows: Sect.~\ref{background} introduces the main background concepts necessary to understanding our proposal and its position in the literature. It includes philosophical theories and related works. 
Sect.~\ref{model} presents the proposed model and its required interfaces. Sect.~\ref{evaluation} evaluates the proposal based on some examples that allow us to identify some limitations and advantages that the model offers on the agent and institutional perspective. Finally, Sect.~\ref{result_and_discussion} presents some conclusions about this work and suggests future works.

\section{Background}
\label{background}

This section presents the main concepts used in our approach.
Subsection~\ref{philosophical_theories} presents the philosophical concepts that support our contribution. Subsection~\ref{institution_MAS} briefly describes the state of the art on artificial institutions with respect of the explicit representation of the functions associated to the institutional concepts.

%Our model is strongly based on some philosophical concepts and theories presented in Subsection~\ref{philosophical_theories}. It is essential that the reader understands the concept of purposes and why they are related to the agents' practical interests. This is the key concept of this work. Finally, in Subsection~\ref{institution_MAS}, some related works that implement artificial institutions are presented, and the differences between these works concerning this one are discussed. 

\subsection{Institutions according to philosophical theories}
\label{philosophical_theories}

An institution is composed of institutional facts~\cite{searle1995construction,searle2010making}. These are based on status-functions and constitutive rules. Status-functions are statuses that have associated functions.
These statuses enable concrete elements to perform functions (associated with the statuses) that cannot be explained through their physical virtues.
Constitutive rules specify the assignment of status-functions to concrete elements with the following formula: \emph{X} count-as \emph{Y} in \emph{C}. For example, a piece of paper count-as money in a bank, where X represents the concrete element, Y the status-function and C the context where that attribute is valid. 
Statuses are imposed on objects when the status-related functions meet some \emph{Purpose}.
The functions are called \emph{agentive functions} because they are assigned from \textit{practical interests of the agents}~\cite[p.20]{searle1995construction}. These practical interests of agents are called \emph{Purposes}. Since the institution is formed by people (i.e., agents) and their collective agreements ~\cite{searle1995construction}, it is possible to say that the agents themselves (through their purposes) assign meaning to the status-functions.
In other words, a purpose is the interpretation of a function performed by an element that carries a status from the agents' perspective.
%For example, when we say ``checkmate" after moving a piece in a chess game, we are putting the purpose of winning the game in the movement of that piece. 
For example, someone has the purpose of winning a chess game when it leads the chessboard to a circumstance that count-as a checkmate.
This purpose does not occur naturally. It is attributed through the practical interests of the agents playing the game (i.e., under that context). Fundamentally, both agents involved in the game must have the same understanding of these facts (i.e., about the function and their purpose). Otherwise, none of them achieve their social goal.
Searle asserts that someone must be \textit{capable of understanding what the thing is for}, or the function could never be assigned~\cite[p.22]{searle1995construction}. Understanding a function requires to understand for what it serves (i.e., its purpose). In the case of chess game, %the purpose of ``checkmate'' 
the purpose of moving the piece to a checkmating position is to win the game. This purpose is in line with the interests of the agents who are playing the game (i.e., its is understood by the people involved in the institution).

While Searle suggests that status-functions are a consequence of collective intentionality, their origin remains, at least in part, unexplained.
For example, throughout history, human societies agreed on assigning the function of money to a piece of paper, a shell or a portion of salt.
However, the function seems not having a genesis. 
It is hard to establish when money or any other social objects were invented.
It is also even more difficult to explain the nature of the collective intentionality that motivates people to act in different ways when they have contact with a concrete element constituted with a status. 
Indeed, Searle considers status-functions as a unique abstraction whose function depends on the individual interpretation of each individual, without worrying about how these functions are represented and shared.

To address the mentioned issues, Ferraris~\cite{condello2018two,condello2019money} proposes to ground the social reality on structures
called Documentality. These structures of documents store informations that not only describe or prescribe, but they actually build social objects. 
Such a structure makes it possible to explain the staying and persistence of functions (and his purposes) associated with status over time. The speech acts that gave origin to functions and status were written and stored in documents that run through time. These documents allow people to learn theses structures through study, perception, etc.
For example, the money exercise its functions in the individual intentions only based in the memory (and consequently in the set of functions) that the social objects recover of the individuals with the base on the recording.

To summarize, social objects are used to externalize the set of recordings that allow that individuals to remember the functionalities that a \emph{status} (e.g., money) makes available by being assigned to a social object (e.g., paper note).
%According to Ferraris, the social object itself, through of the individual intentionality, recovering the records stored in documents that remit the functions that they perform, that determines your value.
From these theories, it is possible to conclude that an additional system of elements is required so that functions and status can persist and have value recognized over time within the social reality~\cite{condello2018two}.
A similar system can be applied to MAS to make explicit the functionalities of the status-functions that compose the institutions.
%Such a system can also bring benefits to the MAS scenario, where the additional system indicate a useful way of understanding the status-functions that composes the institutional reality. 
It will permit to improve the agents' reasoning about the satisfaction of their goals and overcome the difficulties that motivate the realization of this work.

\subsection{Institutions in MAS}
\label{institution_MAS}

The main idea of using artificial institutions as a counterpart of human institutions in computer systems has inspired work in MAS. In different ways, these works use the \textit{count-as} relationship, established through the constitutive rules proposed by Searle, to support the regulation of the system~\cite{Brito2016}. 
This work considers that the constitutive relation, as well as the related concepts (e.g., status-functions) provide functions to be exploited by the agents to achieve their goals
%In this work, the \textit {count-as} relationship is considered, as well as other concepts (e.g., \emph {Status-Function}, norms, etc.), as components of the institutional reality in MAS. 
The purpose of this section is to review state of the art on artificial institutions with respect of the explicit representation of the functions associated to the institutional concepts.

Works on Artificial Institutions are usually inspired by the theory of John Searle \cite{searle1995construction,searle2010making}. Some works present functional approaches, relating brute facts to normative states (e.g., a given action counts as a violation of a norm). These works do not address ontological issues, and, therefore, it becomes even more difficult to support the meaning of abstract concepts present in the institutional reality. Other works have ontological approaches, where brute facts are related to concepts used in the specification of norms (e.g., sending a message counts as a bid in an auction). However, these works have some limitations that are discussed below.

Some approaches allow the agents to reason about the constitutive rules~\cite{cliffe2006specifying,fornara2011specifying,brito2016model,CARDOSO2007,Vigano2008,aldewereld2010making}. However, generally the \emph{status-function} (\emph{Y}) is just a \textit{label} assigned to the concrete element (\emph{X}) and used in the specification of the regulative norms. Therefore, \emph{Y} does not seem to have any other purpose than to serve as a basis for the specification of stable regulative norms~\cite{vazquez2008human,aldewereld2010making}.
Some exceptions are (i)~in the works of \cite{fornara2009ontology,fornara2010representation,fornara2011specifying,fornara2012using} where \emph{Y} represents a class formed with some properties as roles responsible for executing actions, time to execute them, condition for execution, etc.; (ii)~in~\cite{vazquez2008human} where \emph {Y} is a general concept, and \emph {X} is a sub-concept that can be used to explain \emph{Y}. %and, in general, in classified works as functional. 
Although the exceptions contain more information than just a label in the \emph {Y} element, these data are somehow associated with regulative norms.
%the agent until is able to reason and understand what actions can be performed in the environment to satisfy the normative specifications. 
%However, because there is no model that explains the meaning of these actions in the institutional context, the agent may have difficulties in understanding that the actions performed can also satisfy their social objectives.
There are no models that make explicit what the constituted elements (i.e., the status-functions) perform in the institution.
Thus, the agents may not understand that the actions performed can also satisfy their social objectives.
For example, the agent's goal \emph {have a book} is a purpose of the \emph{payment} status-function. Considering the previously described example of selling books, there is currently no way for \emph{Bob} and \emph{Tom} to understand that the actions performed, if interpreted through their \emph{purposes}, can also satisfy their objectives. 
It occurs because there is not work that made explicit the purposes of the status-functions.
If the purposes of status-functions were made explicit, the status-functions' name would be less relevant to the system's correct functioning. Also, new agents could enter the system and understand the purposes of carrying out some functions that have institutional interpretation and thus resolve themselves to satisfy their objectives.

The limitation discussed above indicates the need to develop a model that explains the purposes of status-functions belonging to institutional reality.
Aguilar et al.~\cite{Rodriguez-Aguilar2015} corroborate this conclusion by stating that institutions have not yet considered how to help agents in decision-making, helping them to achieve their own goals.
The modeling of purposes of status-functions, described in the next section, is a step to fill this open gap.

\section{The purpose of status-functions}
\label{model}

Inspired by the document structures suggested by Ferraris~(cf.~Section~\ref{philosophical_theories}), this section describes a model to specify the purposes associated with the status-functions in artificial institutions. The focus is on the main concepts and their relations. %The benefits of implementing the proposed model in a multi-agent system can be seen if the model is implemented in a data structure (for example, an ontology) that allows agents to access the available information and reason about it.  However, the implementation of this model is beyond the scope of this paper. 
The practical implementation of this model is beyond the scope of this paper.

\begin{figure*}[!ht]\centering 
	\includegraphics[width=1.0\linewidth]{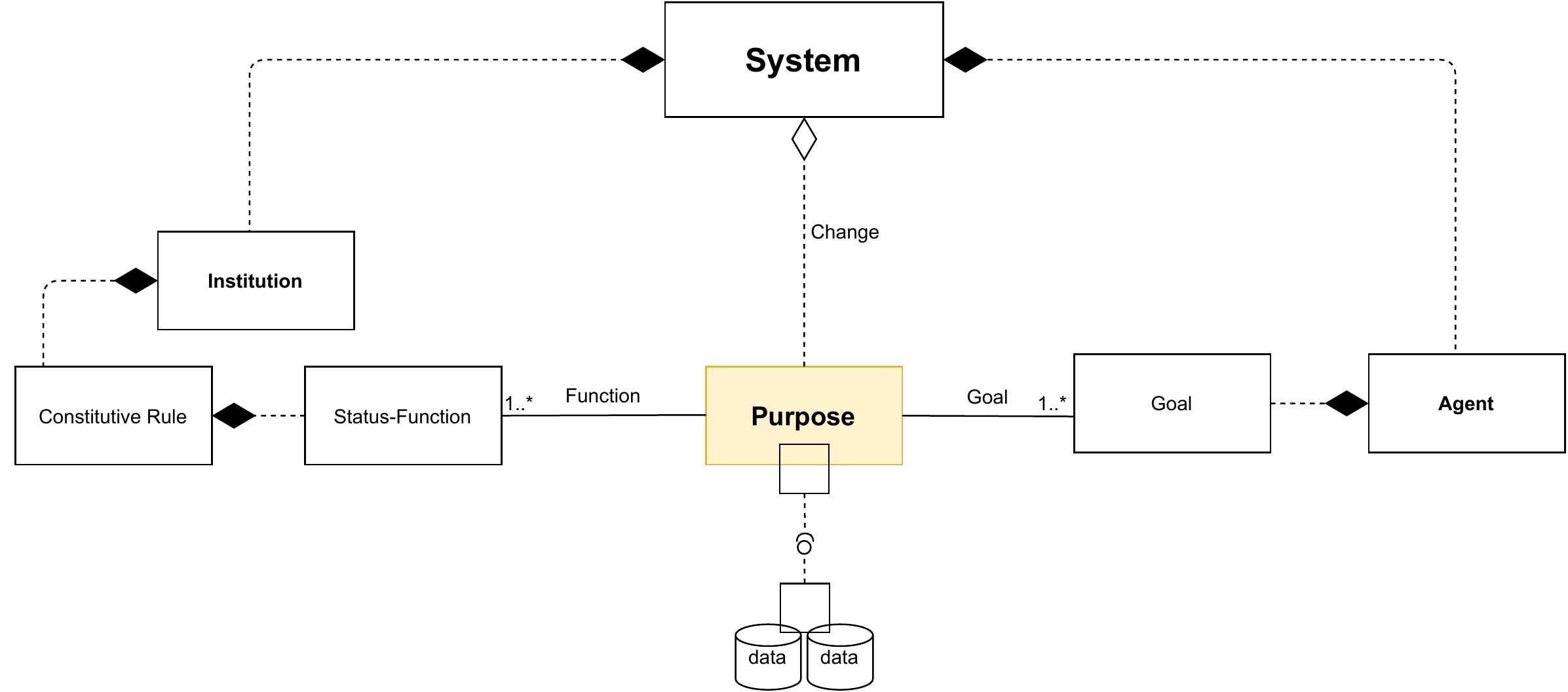}
	\caption{Overview of the model.}
	\label{fig:overview_model}
\end{figure*}

%O principal do meu modelo é o proposito. O que é um proposito. O que um proposito tem.  (Proposito tem um goal e uma SF associado que é detalhado em uma base de conhecimento).

The model proposed in this work is composed of \emph{agents}, \emph{institutions}, and \emph{purposes}.
\emph{Agents} are autonomous entities that pursue their goals in the system~\cite{wooldridge2009introduction}. 
Through MAS definition (cf.~Section~\ref{introduction}), we can see that \emph{goal} is a fundamental concept to understand and program MAS. 
The literature presents several definitions of \emph{goal} that are different but complementary to each other~(see more in~\cite{boissier2020multi,winikoff2002declarative,hindriks2000agent,van2003agent,hubner2006declarative,nigam2006dynamic}). In this work, \emph{goals} are something that agents aim to achieve (e.g.~the holding of a certain state, the performance of an action, etc.). 
%Any multi-agent system (or simply system) is composed of entities called agents. 

\emph{Institutions} provide the social interpretation of the environmental elements of the system. The several models of artificial institutions are usually inspired by the theory of Searle, considering that constitutive rules specify the assignment of status to those elements, enabling them to perform functions in a certain social context. These statuses with their associated functions are then called status-functions. 
The assignment of status-functions to the environmental elements is specified through constitutive rules. 
These rules are generally expressed as \emph{X count-as Y in C} where \emph{X} represents an environmental element (i.e., a brute fact), \emph{Y} represents a status-function to be assigned to X, and \emph{C} represents the context under which the constitution takes place. It is beyond the scope of this paper to propose a model of artificial institution. Rather, it considers this general notion of institution as the entity that constitutes status-functions, that is adopted by several models in the field of MAS.

%\emph{Purposes} are the central concept of the model. 
While \emph{agents} and \emph{institutions} are known concepts, \emph{purposes} are introduced in the proposed model.
The \emph{purposes} are the practical interests of the agents that can be satisfied by the functions associated with the status-functions (cf.~Section~\ref{philosophical_theories}). If a goal is something \emph{desirable} by the agent and a purpose of a status-function expresses practical \emph{interests} of the agents, then we propose to link them. 
In this work, it is assumed that \textit{agents have goals that are satisfied by the purposes of the status-functions}. 
%Therefore, a purpose is a relation between agents' goals and functions of the status-functions. 
In other words, a purpose is the social interpretation of a function linked to a status aligned with the agent's social goal. Therefore, the difference between a purpose and an agent goal is the social characteristic (collective acceptance, etc.) that a purpose requires. Without this social interpretation, a purpose and consequently a social goal aligned with that purpose cannot be achieved.
In addition, it is considered that the purposes are detailed through a knowledge base. 
According to Borst~\cite{borst1999construction}, a knowledge base is an adequate abstraction to represent a shared conceptualization within a context.  Therefore, detailing through knowledge bases is essential for the model to be flexible enough and represent the purpose of status-functions in different contexts.

Shortly, \textit{we are stipulating a relationship between functions of status-functions and purposes and between these purposes and goals of agents}.
Thus, if (i) an agent has a goal that satisfies a purpose, (ii) this purpose is associated to a status-function, and (iii) a constitutive rule specifies how a status-function is constituted, then it is explicit how the agent should act to achieve its goal (i.e., acting to constitute the status-function that is associated to the same purpose as the goal).
In the previous example, Bob can know that he satisfies his social goal consulting the purpose of \emph{payment} status-function and subsequently consulting the institutional specification to understand what concrete action is constituting the \emph{payment} status-function.

\section{Evaluation}
\label{evaluation}

To illustrate the scenario of using institutions in multi-agent systems, consider an open Multi-Agent System where the agents Bob, Alice, François, and João aim to ``have a book". Different programmers have developed the different agents' specifications, that are slightly different from each other. The agent Bob has the goal of \emph{have a book}. Alice has the goal of \emph{get a book}. François has the goal of \emph{obtenir a book} and João has the goal of \emph{ter a book}. 
This example is used in the next subsections to illustrate the limitations of not having an explicit representation of the purposes of the status-functions. Subsection~\ref{agents_limitation} describes the disadvantages of not using the model from the perspective of the agents. Subsection~\ref{agents_model} shows how the model overcomes the disadvantages described in the previous section. The subsection~\ref{institution_limitation} describes the disadvantages of not using the model from the institution's perspective. Finally, the subsection~\ref{institution_model} discusses the advantages that the model offers when used from an institutional perspective.

\subsection{Limitations from the agents' perspective of not using the model proposed in this work}
\label{agents_limitation}

\begin{figure*}[!ht]\centering % Using \begin{figure*} makes the figure take up the entire width of the page
	\includegraphics[width=1.0\linewidth]{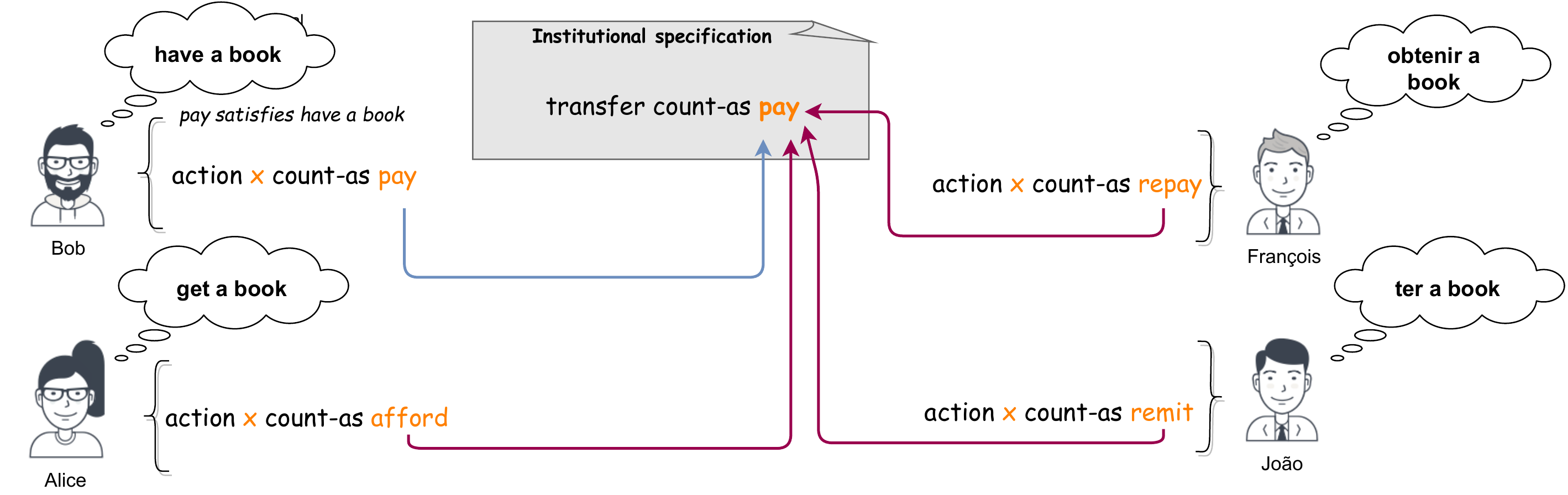}
	\caption{Agents with the same objective operating in the same institution.}
	\label{fig:many_agents_one_institution}
\end{figure*}

In the section, we consider a scenario where all agents described in the introduction are located in a book store~(Figure~\ref{fig:many_agents_one_institution}). This system is instrumented with an institution that contains a constitutive rule stating that the concrete action \emph{transfer count-as pay}. Such a system could include other status-functions and constitutive rules but, for simplicity, we focus only on this case to illustrate the main features of the model proposed~(cf.~Section~\ref{model}).

The first problem observed is the incompatibility of the specification of the agents with the institutional vocabulary. 
For example, consider that \emph{Bob} has a plan to achieve his goal and this plan requires him to take a concrete action \emph{x} that is institutionally interpreted as \emph{pay}. In this case, both specifications (Bob and institution) are compatible because there is a constitutive rule %constituting the
stating that a concrete action x counts as \emph{pay}.
By reading the book store's institutional specification, \emph{Bob} discovers the action that fills his plan (i.e., that \emph{x} is \emph{transfer} in the institution).
However, the problem occurs when other agents have plans considering other words for \emph{pay} (as afford, replay, and remit). Their specifications are not compatible with that of the institution.
\textit{This is because the status-function is just a label}. If the label specified in the agent's plans (e.g., afford, repay, and remit) is different from the label described in the institutional specification, the agents cannot resolve themselves which are the proper action to perform.
%Será necessário falar na possível solução?
%The possible solution to this problem on the agent side is to change their code to be compatible with that of the institution.
%However, in a scenario where there are many agents, this solution may be impractical.
An ad-hoc solution for this problem is to modify the agents' code so that they are compatible with the institution. However, in a scenario where there are many agents, this solution may be impractical.

Another problem observed is that the agents are unable to reason\footnote{Reasoning about the satisfaction of their social goals in the context of this work means that agents consult the purposes (i.e., the consequences of executing a concrete action that has an institutional interpretation) in order to realize that these consequences are analogous to their objectives.} about the satisfaction of their objectives.
For example, the constitutive rule \emph{transfer count-as pay} does not make it explicit that the function of the status-function \emph{pay} can lead to a state of the world where the agents have achieved their goals. Therefore, the agents cannot understand that the transfer action~---~institutionally interpreted as \emph{pay}~---~takes them to a state of the world where they have achieved their objectives.
Institutions, as they are currently conceived, do not provide instruments for agents to reason about actions based on their institutional meaning and, therefore, conclude that they satisfy their social objectives.
The solutions to solve this problem transfer the responsibility of checking if the agent's social objective has been satisfied to the system. However, these solutions decrease the agents' autonomy (they need to follow regulatory rules, perform actions that do not know the consequences, etc.).

%Another problem observed in this scenario is that the constitutive rule \emph{transfer count-as pay} does not make it explicit that the function of the status-function \emph{pay} can lead to a state of the world where agents achieve their goals. Therefore, the agents cannot understand that the transfer action~---~institutionally interpreted as pay~---~took them to a state of the world where they are consistent with their objectives.
%This problem is usually circumvented by shifting responsibility for checking when the agent's goal is satisfied for the system. However, agents are unaware of the satisfaction of their goals.

\subsection{Advantages from the agents' perspective of using the model proposed in this work}
\label{agents_model}

\begin{figure*}[!ht]\centering % Using \begin{figure*} makes the figure take up the entire width of the page
	\includegraphics[width=1.0\linewidth]{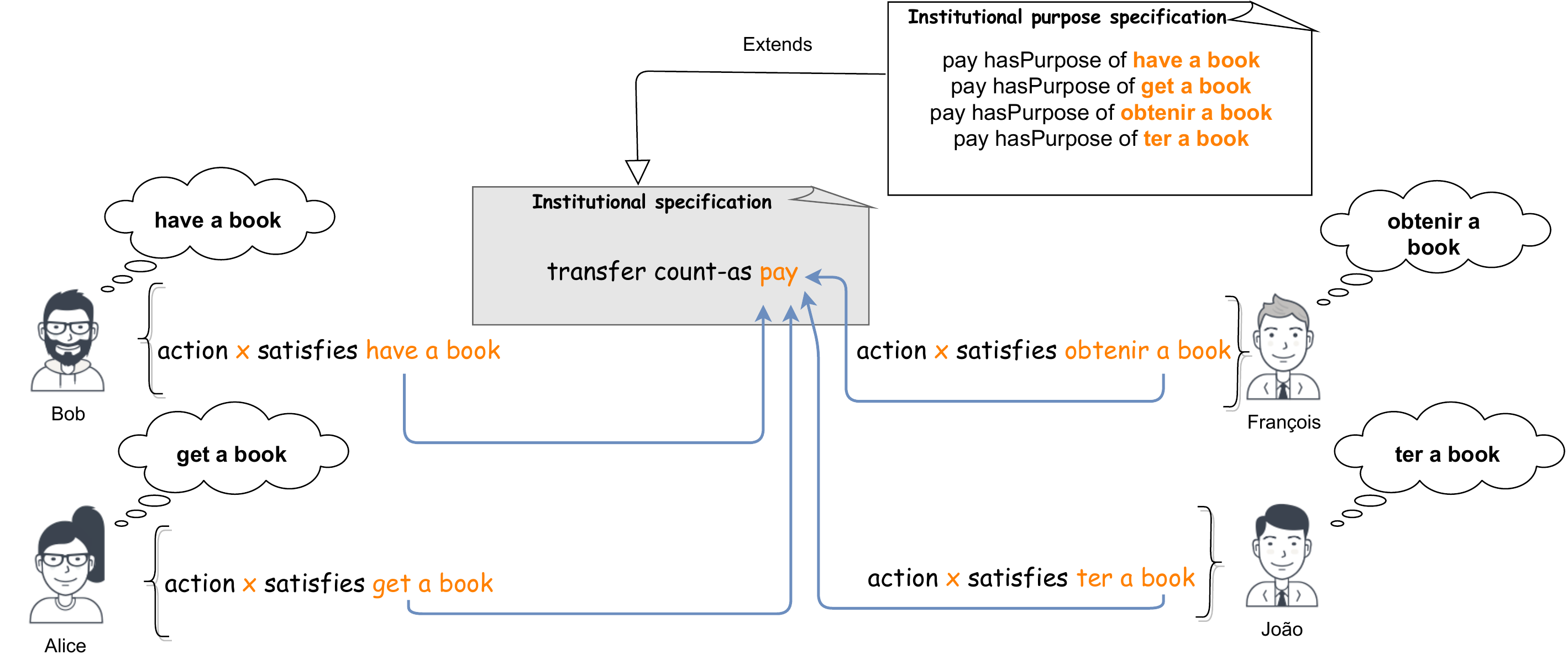}
	\caption{Use of the proposed model in an institutional specification.}
	\label{fig:many_agents_one_institution_solution}
\end{figure*}

To illustrate how the proposed model solves the limitations from the agents' perspective, consider again the example involving Bob, Alice, François, and João and the common social goal of having a book~(Figure~\ref{fig:many_agents_one_institution_solution}). 
However, the purposes related to the status-function \emph{pay} are now explicit.

% Esse "tornar as coisas compatíveis" demanda que se -- desconsidere -- o vocabulário da especificação e se passe a observar o conteúdo por trás disso. Por exemplo, ao invés de Bob ter um plano pra executar a ação X que conte como pay, bob precisa ter um plano pra executar uma ação que interpretada institucionalmente tenha o propósito X. Isso significa que tanto faz se o status for pay, buy, payment, afford, etc. o que importa é o "conteúdo (i.e., o propósito) do status. Essa mensagem precisa ficar "clara" aqui, porque a compatibilidade só pode ser resolvida se houver também uma ligeira mudança no paradigma de se especificar planos nos agentes. A compatibilidade aqui é obtida através da segunda possibilidade que colocamos na nota de rodapé, ou seja, if the programmer encodes compatibility with the institutional specification within the agent. Só que ela é um pouco mais elegante, ou seja, não é hardcoding. 
First, the agent's specification and institutional vocabulary become compatible. This is due to the agents' plans being specified considering the purposes related to the status-functions instead of the nomenclature used in the specifications. For example, in the subsection~\ref{agents_limitation}, Alice's plan requires her to perform an action that counts as \emph{pay}. This could bring some incompatibilities as already discussed. The existence of purposes related to status-functions allows Alice to perform an action that has the purpose of getting a book. The main difference with this proposal is that the institutional nomenclature of status-functions (i.e., \emph{pay}) is irrelevant as long as the agent has access to the status-function's purposes. In other words, the agent's plan will not look to the status but to the purposes associated with it.
Agents can act in the institution according their purposes without requiring changes in their code.

%In other words, Bob executes the \emph{transfer} action that is institutionally interpreted as \emph{pay} and has the purpose of \emph{having a book} that is precisely his objective. Based on this mechanism, the nomenclature of agents does not need to be aligned with that of the institution and, or, agents do not need to contain codes that make them compatible with the institution.  (ii) Agents can understand the consequences (i.e., the purposes) of the status-function functions assigned in concrete actions. This allows them to understand when the actions are taken (interpreted institutionally) satisfy their goals or when the actions are in disagreement with their beliefs. For example, if the transfer action institutionally interpreted as pay did not have the purpose of having a book, Bob would be able to understand this and decide whether or not to execute it.

Another advantage is the improvement in the agents reasoning about the satisfaction of their goals. By adding purposes related to status-functions, agents can reason about the purposes (i.e., the consequences) of carrying out actions constituting status-functions. From this, the agents can check which actions are aligned to their social goals and, if performed, these actions can satisfy them.
For example, the status-functions \emph{pay} has some associated purposes, among them, that of \emph{having a book}. If \emph{Bob} looks for this own social goal of having a book in the purposes, he can infer that, the status-function \emph{pay} is of his interest and, moreover, that the concrete action of \emph{transfer} is counting as \emph{pay} in this institution. The very \emph{x} action for his plan in this institution is thus \emph{transfer}.
This solution allows the agent himself to check if his social objective can be satisfied. This solution seems to be appropriate because it prevents the agent from following unnecessary regulative rules or performing actions he does not know the consequences.

\subsection{Limitations from the institution' perspective of not using the model proposed in this work}
\label{institution_limitation}

%Third, the same institution may not be compatible with different agents with similar objectives but coded with other terms. This is the same limitation as to the previous example but seen from the institution's perspective. The institution needs to add new status-functions compatible with agents' specifications to its specification to get around this problem. However, some status-functions can make it inconsistent. For example, it is not possible to specify the \emph{payment} status-function in a library that only lends books.

\begin{figure*}[!ht]\centering % Using \begin{figure*} makes the figure take up the entire width of the page
	\includegraphics[width=1.0\linewidth]{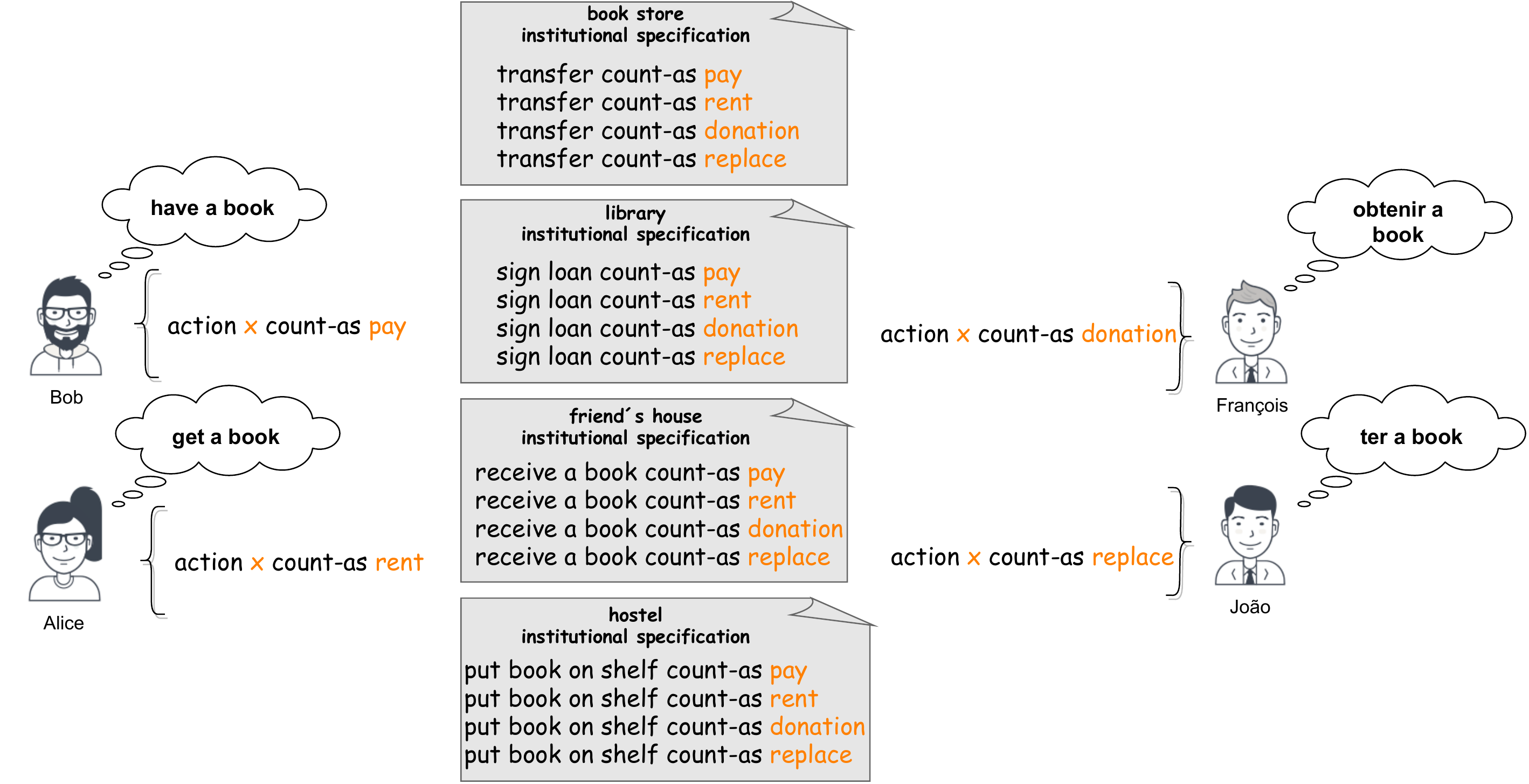}
	%\caption{Agents with the same objective operating in different institutions.}
	\caption{The institution prepared for different agents.}
	\label{fig:many_agents_many_institutions}
\end{figure*}

Consider that Bob, Alice, François and João still have the same social goals but can move through four different institutions. They are (i) a book store, (ii) a library, (iii) a bookshelf at a friend's house, and (iv) a hostel shelf~(Figure~\ref{fig:many_agents_many_institutions}).
The first institution contains a constitutive rule stating that the action \emph{transfer count-as pay, rent, donation and replace}.
The second institution contains a constitutive rule stating that \emph{signing a loan count-as pay, rent, donation and replace}. The third institution contains a constitutive rule stating that \emph{receiving the book from a friend count-as pay, rent, donation and replace}. Finally, the fourth institution contains a constitutive rule stating that \emph{putting a book on a shelf count-as pay, rent, donation and replace}. All these scenarios are part of a similar context. That is, it involves obtaining books.

The problem from an institutional perspective is that institutions have to be specified for every possible incoming agent. In the example, they have to have four constitutive rules, one for each agent. For every new different agent, a new constitutive rule have to be added in each institution. The institutional specification is thus quite dependent of how agents are programmed. The institutional developer has to be worried about the agents' internals instead of specifying a good institution (independent of the incoming agents).

In fact, there are some solutions to work around this problem. The first one is to
change the institution's specification by adding new constitutive rules. However, it cannot be guaranteed that the new constitutive rule makes sense in that system or institution (which sometimes makes this solution unfeasible). A second possible solution is to
change the agent's code. However, as already mentioned in~\ref{agents_limitation} subsection, this solution becomes impractical if many agents are operating on the system.
Furthermore, if a new system is available for agents or new agents to join different systems, new changes need to happen. As it is currently conceived, the institution is not open enough to support the performance of different agents designed with different (but similar) plans and objectives.

\subsection{Advantages from the institution' perspective of using the model proposed in this work}
\label{institution_model}

\begin{figure*}[!ht]\centering % Using \begin{figure*} makes the figure take up the entire width of the page
	\includegraphics[width=1.0\linewidth]{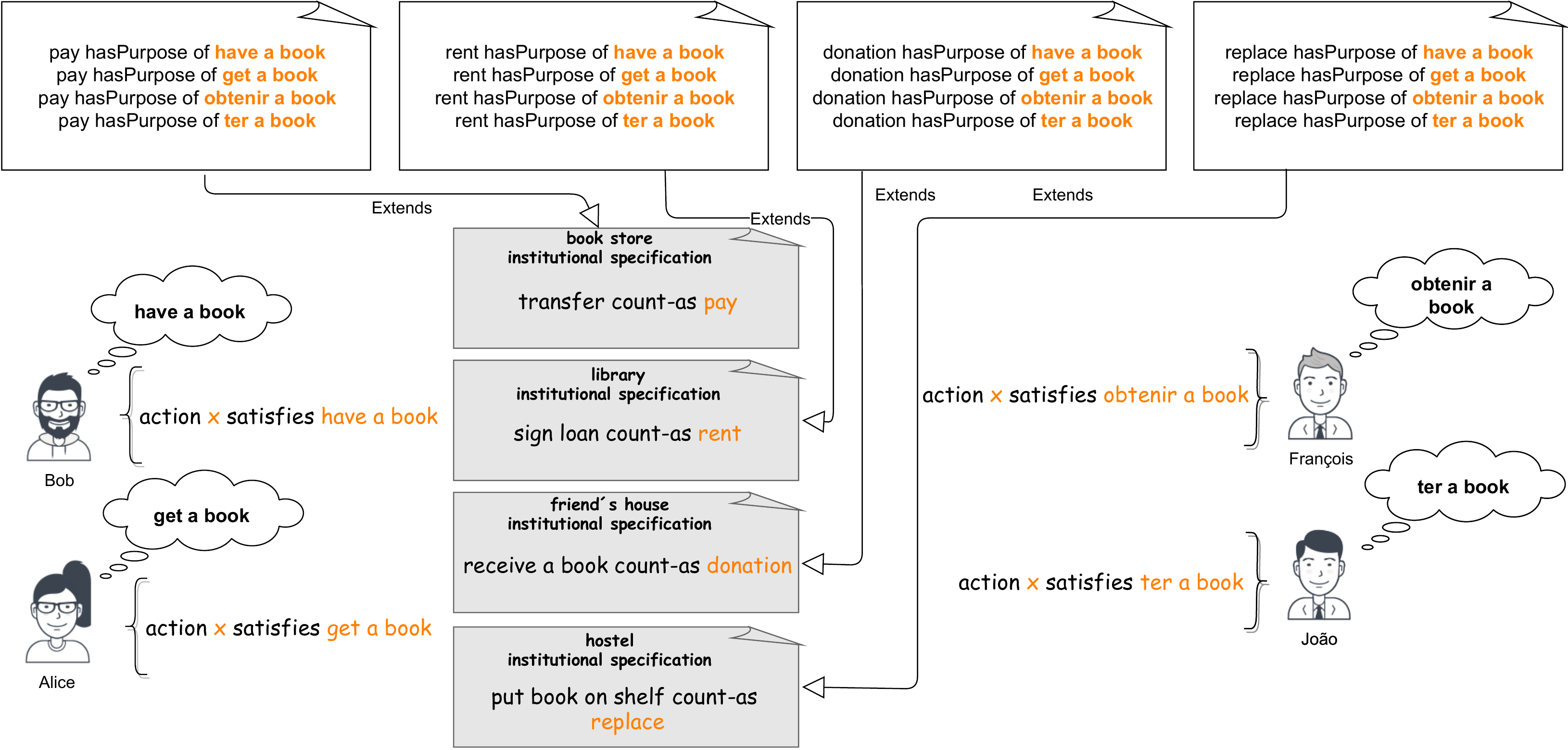}
	\caption{Use of the proposed model in some institutional specifications.}
	\label{fig:many_agents_many_institutions_solution}
\end{figure*}

To illustrate the proposed model this model's use from the institution perspective, consider again the example involving Bob, Alice, François, and João still have the same goals but can move through four institutions containing four different institutional specification ~(Figure~\ref{fig:many_agents_many_institutions_solution}). 
In addition, there are purposes related to the status-functions.
Each institution contains its own institutional specification that is extended by adding purposes related to the status-functions. These purposes represent the interests of agents possibly operating in the institution.

We can notice that the institutional specification, in particular the constitutive rules, is not conceived for the different possible incoming agent. The link between the status-functions and agents' goals is done in a new document properly conceived for that. If a new agent is considered to be participant of the institution, only this document has to be adapted. The constitution is kept unchanged.
For example, the \emph{book store} specification contains an associated document that describe the purposes of the \emph{pay} status-function. In this case, \emph{pay} is related to the social goals (i.e., has the purposes) of \emph{Bob}, \emph{Alice}, \emph{François} and \emph{João}.

\section{Results and discussions}
\label{result_and_discussion}
 
%maiquelb: Acho que além disso, há outro problema (talvez possamos conversar sobre na reunião): Nesse caso, o sistema não é totalmente aberto à participação de qualquer agente, porque, para que outros agentes participem, precisam ser recodificados para serem compatíveis com a instituição.

The problem motivating this paper is the impossibility of the agents to reason about the functions associated with status-functions representing the institutional interpretation of certain facts that occur in the environment.
This problem is partially solved by computational models that implement artificial institutions. 
%and interpret certain facts about the environment from an institutional perspective.
However, these models do not represent the purpose of this interpretation from an agent perspective. Agents have to be hard-coded to know which status-functions can be useful for them to achieve their social goals and institutions must be recoded to receive new agents specified by different parties. Considering this problem, we propose a model to express the purposes associated with the status-functions that compose the artificial institutions. % presents on the MAS.
The conception of our model is an adaptation, from a particular point of view, of Searle's and Ferraris's  theories~\cite{searle1995construction,searle2010making,condello2019money,condello2018two} which claims that the purposes are specified by practical interests of the individuals that can be satisfied by the functions associated to the status-functions (cf.~Section~\ref{philosophical_theories}). 
We assume that the purposes are defined based on the interests of the agents. However, they are specified by a party involved in the development of the system. In other words, it is not the agents themselves who define the purposes based on collective agreements, etc.

There are some advantages of such conception that we discuss below.
The first one is related to the \emph{flexibility} of
the agents, that can achieve their social goals even being developed by different programmers
For example, the subsection~\ref{agents_limitation} illustrates a scenario where all agents are located in the same institution (e.g., book store). 
In this example, specified agents with vocabularies other than the institutional vocabulary may have difficulties to (i)~execute their plans and (ii)~understand when their social goals are reached.
The solution to this problem is described in subsection~\ref{agents_model} where the agents' plans can be specified considering the purposes related to the status-functions rather than the status nomenclature.
The advantages are (a) the agents can reason about these functions and adapt in different scenarios to satisfy their plans and (b) by reasoning about the functions of the status-functions, the agent can perceive that these functions are similar to their interests and therefore they can help the agents to reach their social goals.
The agent's capability to reason about the functions and adapt to different scenarios is an important advance in open systems' flexibility~\cite{aldewereld2010operetta,zambonelli2000organisational}.
The agent's understanding of what makes his social objective satisfied is also an important advance in his autonomy~\cite{Rodriguez-Aguilar2015}. In this case, the agent can reason about the actions in the plans and the regulative rules that govern the system. In both cases, the agent has greater autonomy and flexibility in deciding whether a particular action will help him reach his social goal.

The second advantage is related to the \emph{institution's flexibility} in being prepared to receive different agents designed by different developers.
For example, the subsection~\ref{institution_limitation} illustrates a scenario where all agents can enter and leave different institutions. In this example, institutions are not prepared to receive different agents. Then, the developer needs to modify the institutional specifications. %or wait for agents to modify their internal codes to become compatible with the institutional specification. 
However, in open systems, it is not possible to predict which agents will enter or leave the institution~\cite{fornara2004agent,Piunti2009}. 
The solution to the problem is described in subsection~\ref{institution_model}: an the explicit link between status-functions and purposes that allows us to prepare the institution for new agents. 
%the model allows making explicit purposes related to the status-functions of the institutional specifications.
%The advantage is that the model allows specifying the purpose of status-functions in the contexts in which the system and the institution operate in a more precise and generic conceptual vocabulary (through ontologies), making the institution more flexible and stable.

The advantages can also be seen in the concepts of \emph{cohesion} and \emph{coupling} present in software engineering. The cohesion of a module corresponds to the degree to which the module is dedicated to implementing only one responsibility of the system~\cite{pfleeger2010software}. Cohesive modules are those that have few responsibilities. This way, maintenance is simpler and avoids side effects. It is easier to change one part of the application without affecting other parts.
The coupling means how much a module depends on the other module to work~\cite{pressman2016engenharia}.
When there is low coupling, the application becomes more flexible, reusable and more organized.
Therefore, the ideal for a system to be flexible and understandable are modules with high cohesion and low coupling~\cite{pressman2016engenharia}. The proposed model has both properties.
High cohesion occurs because the model proposes to extend only the status-functions present in the institutional reality through the introduction of the concept of purposes related to these functions. The low coupling is guaranteed because the model does not depend on the other system modules (environment, institution, agents, etc.) to function correctly. This separation of responsibilities also ensures that the model is flexible and requires less effort to perform maintenance.

According to~\cite{DeBrito2014}, human societies usually have a representation (i) internal of the social reality, i.e., the individuals do not necessarily reason in terms of status-functions, norms, purposes, etc and (ii) implicit, as it is built on top of people’s mental states (that believe, for instance, that a certain man is the king). In the proposed model, purposes compose a system (i) explicit, as it is properly specified through institutional concepts and (ii) external, as it is persisted outside the agents mind.
Such conception is in agreement with some authors that point that institutions can (or perhaps even should) be used for purposes that are beyond the normative ones~\cite{Rodriguez-Aguilar2015,fornara2011specifying,telang2019coupled,Tomic2018,Murray-Rust2015,Padget2018}. In summary, our work proposes an interface to make different couplings in different institutions without changing the institutional specification or the coding of the agents. In other words, agents and institutions remain unchanged because the purpose is the link between agents' interests and the institutions. From this, the change from one scenario to another occurs only in the mapping between the institutional specification and the purposes that reflect the agents' interests.

%This work approaches in a conceptual way the advantages that the model occurs when used. 
As future work, we plan to explore additional theoretical aspects related to the model, such as (i) investigations about how other proposed institutional abstractions fit on the model, (ii) the verification of the consistency among status-functions' purposes and agents' social goals, and (iii) check if the functions related to status must be further detailed. We plan to also address more practical points such as (i) the modeling of a status-functions' purposes based on a real scenario, (ii) the implementation of this model in a computer system and (iii) the integration this model in an computational model that implements the constitution of status-functions and after in an MAS platform.

\bibliographystyle{splncs04}
\bibliography{mybibliography}

\end{document}